\documentclass[apj]{emulateapj}
\usepackage{pslatex}
\usepackage{graphicx}
\shorttitle{X-ray emission in galaxies}
\shortauthors{Chatterjee et al.}
\begin{document}
\title{{X-ray Emission in Non-AGN Galaxies at z $\simeq$ 1}} 
\author{Suchetana Chatterjee$^{1,2}$, Jeffrey A.\ Newman$^{3,4}$, Tesla Jeltema$^{5}$,  Adam D.\ Myers$^{1}$, James Aird$^{6}$, Kevin Bundy$^{7}$, Christopher Conselice$^{8}$, Michael Cooper$^{9}$, Elise Laird$^{10}$, Kirpal Nandra$^{11}$, Christopher Willmer$^{12}$ }
\affiliation{$^{1}${Department of Physics and Astronomy, University of Wyoming, Laramie, WY 82072 USA}\\
$^{2}${Department of Physics, Presidency University, Kolkata, 700073, India}\\
$^{3}${Department of Physics and Astronomy, University of Pittsburgh, Pittsburgh, PA 15260 USA}\\
$^{4}${PITT-PACC, University of Pittsburgh, Pittsburgh, PA 15260, USA}\\
$^{5}${Department of Physics, University of California, Santa Cruz, CA 95064 USA}\\
$^{6}${Department of Physics, Durham University, Durham DH13LE, UK}\\
$^{7}${Institute for the Physics and Mathematics of the Universe, Kashiwa, 277-8583, Japan}\\
$^{8}${Centre for Astronomy and Particle Physics, University Park, Nottingham, NG7 2RD, UK}\\
$^{9}${Department of Physics and Astronomy, University of California, Irvine, CA 92697 USA}\\
$^{10}${Astrophysics Group, Imperial College London, Blackett Laboratory, Prince Consort Road, London SW7 2AZ, UK} \\
$^{11}${Max Planck  Institut f\"{u}r Extraterrestrische  Physik, Giessenbachstra\ss e, 85748 Garching, Germany}\\
$^{12}${Steward Observatory, University of Arizona, Tucson, AZ, 85721, USA}\\
}

\begin{abstract}
Using data from the DEEP2 galaxy redshift survey and the All Wavelength Extended Groth Strip International Survey we obtain stacked X-ray maps of galaxies at $0.7 \leq z \leq 1.0$ as a function of stellar mass. We compute the total X-ray counts of these galaxies and show that in the soft band (0.5--2\,kev) there exists a significant correlation between galaxy X-ray counts and stellar mass at these redshifts. The best-fit relation between X-ray counts and stellar mass can be characterized by a power law with a slope of $0.58 \pm 0.1$. We do not find any correlation between stellar mass and X-ray luminosities in the hard (2--7\,kev) and ultra-hard (4--7\,kev) bands. The derived hardness ratios of our galaxies suggest that the X-ray emission is degenerate between two spectral models, namely point-like power-law emission and extended plasma emission in the interstellar medium. This is similar to what has been observed in low redshift galaxies. Using a simple spectral model where half of the emission comes from power-law sources and the other half from the extended hot halo we derive the X-ray luminosities of our galaxies. The soft X-ray luminosities of our galaxies lie in the range $10^{39}$--$8\times 10^{40}$\,ergs\,s$^{-1}$. Dividing our galaxy sample by the criteria $U-B > 1$, we find no evidence that our results for X-ray scaling relations depend on optical color.   
\end{abstract}

\section{Introduction}

The first systematic studies of X-ray emission in normal galaxies, conducted using the Einstein Observatory, revealed extended and complex X-ray structure in nearby galaxies \citep[e.g.,][]{fabbiano89}. Subsequent X-ray observations at higher angular resolution with ROSAT augmented these studies \citep[e.g.,][]{tanakaetal94,matsuhitaetal94}, but the observations of the {\it Chandra} and {\it XMM}-Newton satellites have revolutionized our knowledge of the X-ray source populations within galaxies (\citealt{o&w09, rovilosetal09}, and \citealt {fabbiano12} for a review). The main components contributing to the X-ray emission in galaxies can be broadly classified into two types---namely, diffuse emission from the hot interstellar medium (ISM) which has a cosmological origin, and pointed emission from astrophysical objects within galaxies such as X-ray binaries, evolved stellar populations, and planetary nebula \citep[e.g.,][]{m&b03, brassingtonetal08, brassingtonetal09, b&p09, fabbiano12}. 

The existence of hot gaseous halos in galaxies is an integral part of the theory of galaxy formation \citep[e.g.,][]{w&r78}. Gaseous halos should be formed when gas condenses on to dark matter halos and undergoes subsequent shock heating and adiabatic compression. Based on this idea, the existence of extended soft X-ray emission around galaxies was predicted by \citet{w&f91}. This established the foundation for interpreting galaxy evolution within the Cold Dark Matter paradigm. Hot gas halos have been seen in early type galaxies since the advent of the Einstein satellite \citep[e.g.,][]{formanetal85, t&f85,canizaresetal87} and work on the phenomenon has been extended by many authors using data from ROSAT and {\it Chandra} \citep[e.g.,][]{osullivanetal01,davidetal06,sunetal07,jeltemaetal08, sunetal09, andersonetal13}. The detection and characterization of hot halos in galaxies has firmly established the basic paradigm that galaxies form within collapsing gaseous halos. 

\begin{figure*}[t]
\begin{center}
\begin{tabular}{c}
        \resizebox{16cm}{!}{\includegraphics{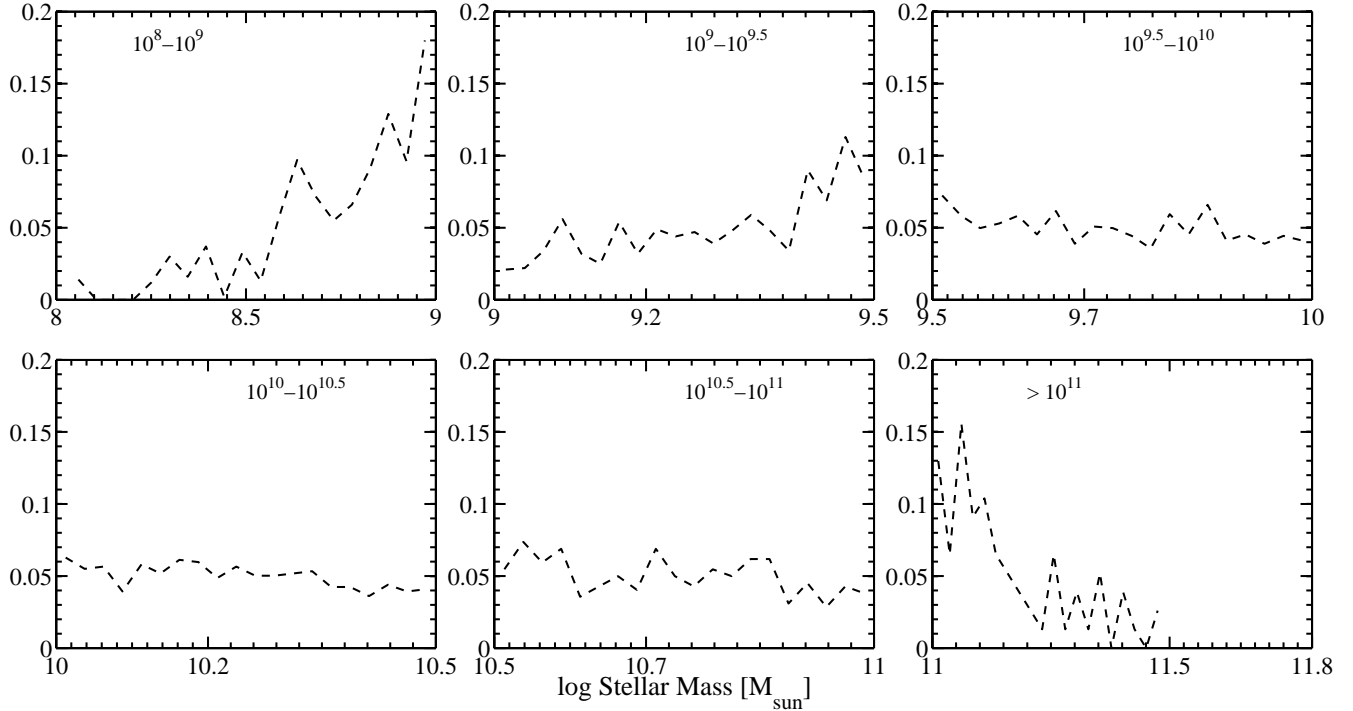}}
      \end{tabular}
 \caption{Distribution of stellar mass in each stellar mass bin. The redshifts of these galaxies are 0.7 $\leq z \leq 1.0$. The redshift distributions are shown in Fig.\ 2. Any galaxy that has been identified as an X-ray bright AGN in the AEGIS-X catalog (L09) has been discarded from our analysis, since we want to minimize the X-ray nuclear emission from AGN.}
\end{center}
\end{figure*}

The origin of X-ray emission from astrophysical sources within galaxies has mainly been studied through their correlation with galaxy stellar luminosities \citep[e.g.,][]{formanetal85, t&f85, canizaresetal87, sarazinetal00, k&f03, sivakoffetal03}. Correlations between X-ray properties and stellar luminosity corroborate the connection between galactic X-ray emission and the stellar component in galaxies. The relative fraction of X-ray emission from hot gas in the halo to that of the stellar component in the galaxy---in order to separate astrophysical signals from cosmological signals---has been addressed in local galaxies with {\it Chandra} \citep[e.g.,][]{fabbiano06}. We note that, in the context of X-ray star-forming galaxies, the diffuse emission is often referred 
to the X-ray emission in the ISM which has been heated by supernovae \citep[e.g.,][]{stricklandetal02, ranallietal08, mineoetal12}. We clarify that in the current paper, the diffuse emission refers to the ISM gas that is shock-heated due to its infall into the gravitational potential well of the galaxy. 

The results show that the relative fraction of X-ray emission from hot halo and stellar components depends on the luminosity and morphology of the galaxy (see \citealt {fabbiano06} and references therein). Further correlations appear to exist between the X-ray gas around galaxies and galaxy environments on large scales \citep[e.g.,][]{ w&s91, b&b00, osullivanetal01, helsdonetal01, e&o06,  m&j10}. However all of these studies have been carried out for samples of local galaxies. To check whether the X-ray properties of local galaxies are ubiquitous features of galaxy evolution, it is important to search for similar correlations at higher redshift. 

In this paper we present a study of the relationship between X-ray luminosity and stellar mass in galaxies approaching a redshift of $z\sim 1$. Our goal is to characterize the X-ray emission from galactic sources and compare them with the results obtained for local galaxies \citep[e.g.,][]{borosonetal11, andersonetal13}. We perform a stacking analysis of galaxies at $0.7\leq z \leq 1.0$, observed in the DEEP2 survey \citep{davisetal03, newmanetal13} by using X-ray data from the All Wavelength Extended Groth Strip International Survey (AEGIS; \citealt{davisetal07, nandraetal07}). We then compute the X-ray properties of these galaxies from the stacked map as a function of galaxy stellar mass and statistically evaluate our findings in light of local galaxies. AEGIS provides a unique combination of wide area and substantial depth, and DEEP2 provides a high sampling rate of spectroscopy. In combination, these factors allow studies of the sources of X-ray emission in galaxies to be extended to higher redshift. 

Our paper is organized as follows. In \S 2 we give a brief description of our data sets and describe our methodology. We present and discuss our results in \S 3 and \S 4 respectively. We finally summarize our conclusions in \S 5.  Throughout the paper we assume a spatially flat, $\Lambda$CDM cosmology: $\Omega_{m}=0.28$, $\Omega_{\Lambda}=0.72$, $\Omega_{b}=0.04$, and h$=0.71$ \citep{spergeletal07}.    
 
\begin{figure*}[t]
\begin{center}
\begin{tabular}{c}
        \resizebox{16cm}{!}{\includegraphics{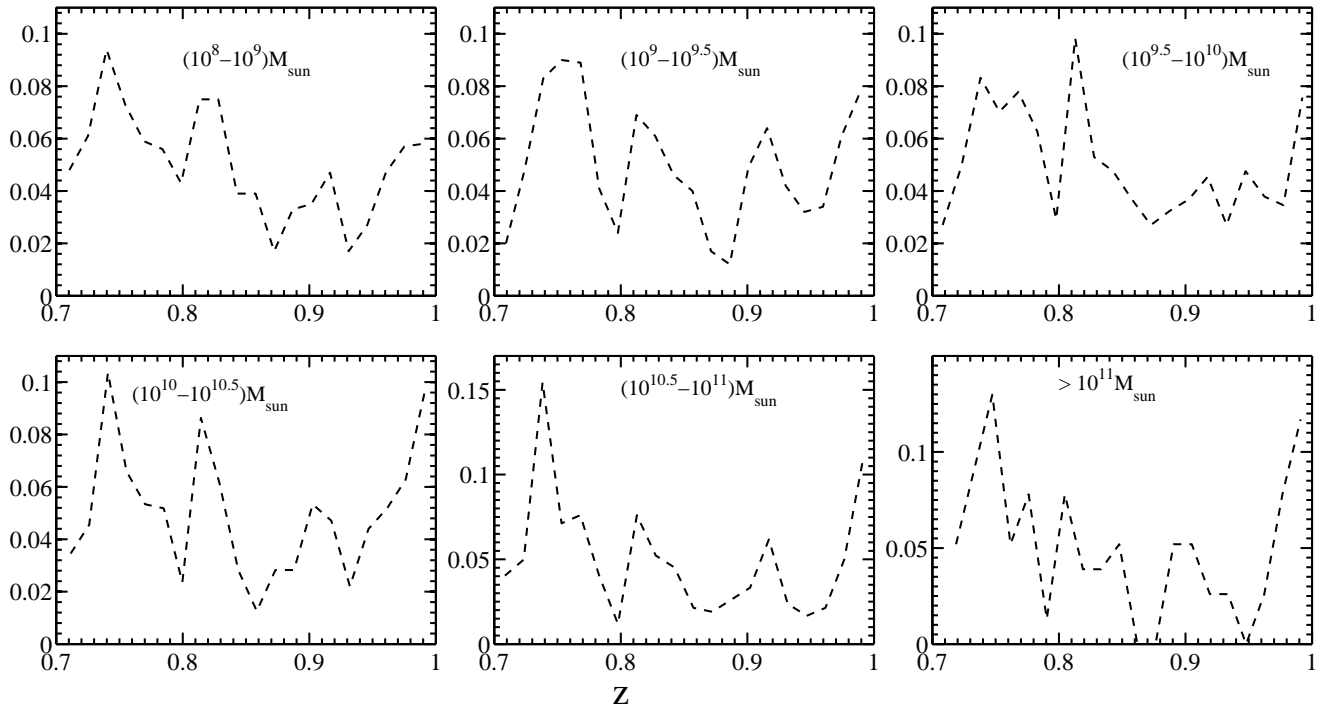}}
      \end{tabular}
 \caption{Distribution of redshift in each stellar mass bin. \S 2 details the construction of samples with matched redshift distributions. The stellar masses are shown in each panel.}
\end{center}
\end{figure*}
\section{Data Sets}

We use data from the AEGIS-X survey (\citealt{lairdetal09}; L09 hereafter) and the DEEP2 galaxy redshift survey. In this section we will provide a brief description of our data sets.

The DEEP2 Redshift Survey (Deep Evolutionary Exploratory Probe; \citealt{davisetal03}) is a project that has measured redshifts of $50{,}000$ galaxies to a limiting magnitude of $R_{AB} = 24.1$ \citep{coiletal04, faberetal07}. The DEEP2 Galaxy Redshift Survey is a spectroscopic survey of galaxies using the DEIMOS spectrograph \citep{faberetal03} on the 10\,m Keck II telescope to survey optical galaxies at $z \sim 1$ in a comoving volume of approximately ${\rm 5 \times 10^{6}\,h^{-3}\, Mpc^{3}}$. The survey covers $3\,{\rm deg}^{2}$, over four widely separated fields to mitigate cosmic variance. Due to the high resolution (R $\sim 5000$) of the DEEP2 spectra, redshift errors, are $\sim 30\,{\rm km\,s}^{-1}$. For more details of DEEP2 observations, catalog construction, and data reduction see \citet{davisetal03,davisetal05,coiletal04a}, and \citet{faberetal07}. 

To measure accurate stellar masses, it is necessary to have photometry extending from the visible into the near-infrared (4,000--22,000 \AA). However, since only $\sim$ 33\% of DEEP2 galaxies with redshifts have stellar mass measurements from $K$ band magnitudes \citep{bundyetal06, conseliceetal08}, we use a different method to measure the stellar masses of the remaining galaxies.

We used the method proposed by \citet{linletal07} and more recently by \citet{weineretal09} to estimate stellar masses for the remainder of our galaxy sample. The rest frame $M_{{\rm B}}$ and $U-B$ color\footnote{We obtain rest frame ($U$-$B$) colors using the method of \citet{willmeretal06}.} are used to estimate the stellar masses using color--M/L relations \citep{b&deJong01, belletal03} and then corrected using $K$-band photometry and SED fitting by \citet{bundyetal06}. We have $14121$ galaxies in the Extended Groth Strip (EGS) field that have measurements of stellar mass. We apply a redshift cut ($0.7 \le z \le 1.0$) to our parent galaxy sample. We also discard galaxies from our sample that are identified as X-ray bright AGN in L09. Our final galaxy sample comprises 2914 galaxies. We subdivide our galaxy sample into six stellar mass bins shown in Fig.\ 1. 

In Fig.\ 2 we present the redshift distribution of our sources. We expect that galaxy X-ray properties will depend on redshift---since galaxy ages, luminosity distance, angular diameter distance, and surface brightness dimming will all evolve with redshift. Hence, to compare the luminosity dependence with galaxy stellar mass we require the redshift distributions to be identical in each stellar mass bin (shown in Fig.\ 1). We note that the redshift distribution of the two lowest stellar mass bins (top-left and top-middle panels in Fig.\ 2; the figure actually shows the matched distribution described below) differ from the other four mass bins. 

To ensure that we can study low and high stellar mass samples over a similar volume, we applied a redshift matching criterion. To obtain matched redshift distributions we stack some galaxies multiple times for the lower stellar mass bins. We note that a unique sample of galaxies do not reproduce the redshift distribution in the two lowest stellar mass bins. Thus, it is necessary to repeat objects in order to match the distribution of our samples. The final redshift distributions for each stellar mass bin are shown in Fig.\ 2. We note that this technique could be problematic in cases where a single object may come to dominate the properties of the stellar mass bin (e.g., if it is the brightest).   

The AEGIS-X survey consists of 8 deep {\it Chandra} ACIS-I fields, each with a total integration time of about 200\,ks. The total area covered in the survey is about $0.67\,{\rm deg^{2}}$. The images are constructed in four energy bands namely, 0.5--7.0\,keV (full), 0.5--2.0\,keV (soft), 2.0--7.0\,keV (hard) and 4.0--7.0\,keV (ultra-hard). The limiting fluxes are $2.37 \times 10^{-16}$, $5.31  \times 10^{-17}$,  $3.76 \times  10^{-16}$, and  $\rm  6.24 \times 10^{-16}  \,  erg \,  s^{-1}  \,  cm^{-2}$, respectively. The limiting flux is defined as the flux to which at least $1\%$ of the survey area is sensitive. \footnote{See L09 for more extensive details of how the data were reduced} A point source catalog in the EGS field has been provided in L09. The catalog consists of a total of $1325$ sources. We identified X-ray bright point sources in our galaxy catalog from this parent catalog. We also used the AEGIS-X point source catalog to mask out all X-ray bright point sources in the EGS field. Note that we also discarded X-ray bright AGN from our galaxy sample. The X-ray bright AGN were identified from the L09 point source catalog. We used a luminosity cut (L$_{X} \ge 10^{41} ergs^{-1}$) to identify the X-ray bright AGN from our source catalog.

\begin{figure*}[t]
\begin{center}
\begin{tabular}{c}
        \resizebox{16cm}{!}{\includegraphics{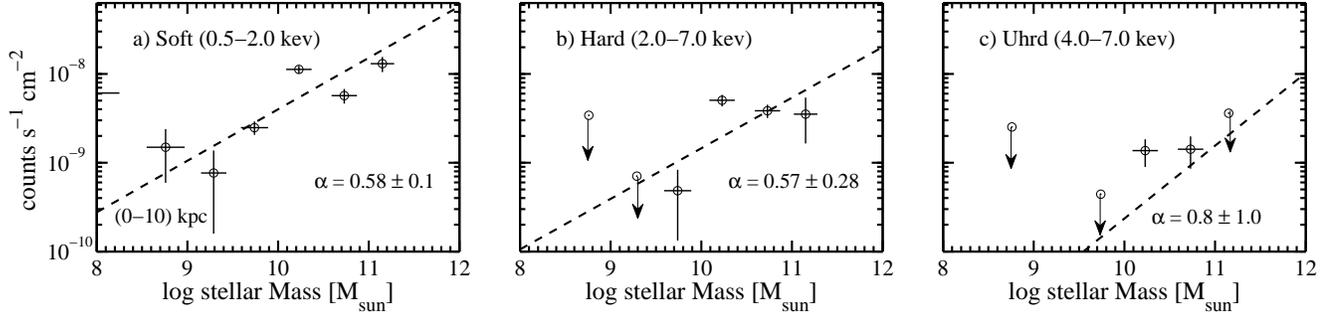}}
 \end{tabular}
 \caption{ Correlation between stellar mass and X-ray counts in the three energy bands. The left-hand, middle, and right-hand panels represent the soft, hard and ultra-hard bands. The data points with error bars are detections whereas the open circles (with arrows) are $2 \sigma$ upper limits. The dashed line in each panel represents the best-fit power law. The best-fit slopes ($\alpha={\rm dlog}N/{\rm dlog}M_{\star}$: $N$ and $M_{\star}$ being X-ray luminosity and stellar mass respectively) are shown in the each panel. When converted to X-ray luminosity (Fig.\ 5), the best-fit slope $\alpha={\rm dlog}L/{\rm dlog}M_{\star}$ is exactly identical to what has been observed with photon counts. We have included both the significant and the insignificant detections for obtaining our best-fits. We note that the slopes for the soft and the hard bands are similar while the best-fit slope for the ultra-hard band is consistent with zero. We also note that the correlation co-efficient is only significant for the soft band. The errors on the stellar masses are the 1 $\sigma$ spread in the distributions and are identical in each panel. See \S 4 for more details.}
\end{center}
\end{figure*}

\subsection{Methodology}
We perform a stacking analysis to identify the X-ray emission, from our galaxy sources. The event files, the exposure time maps and the effective area maps are provided in L09. We first correct our event files using the exposure time-effective area maps. We call these effective area-exposure corrected (EAEC hereafter) maps. The gaps and singularities in the EAEC maps are corrected by assigning zero counts to those pixels where the EAEC maps have singularities (due to zero exposure and/or effective area). We then identify sources in the EAEC maps and select a $5\times 5$ square arcminute region around each source and sum them to construct the stacked image. 

We identify the point sources in each map from the point source catalog presented in L09. We calculate the point spread function (PSF) for each point source using the technique described in L09. The {\it getpsf} routine, provided by L09, is used to obtain the PSFs. We then mask the point sources using the ellipse corresponding to the $95\%$ encircled energy radius (EER). We mask the region which encompasses a circular area with a radius $1.5$ times that of the semi-major axis of the $95\%$ encircled energy ellipse of a particular point source. This allows us to create a conservative mask for each point source. 

We also construct a stacked mask map. It is important to note that a source in one map could be a potential contaminant point source (to be masked) in another map. The EAEC maps are divided by the stacked mask maps to obtain the average photon counts in each pixel of our final stacked maps. 
We construct a total of 18 maps for the three bands. In each band we have six stacked maps corresponding to the six stellar mass bins shown in Fig.\ 1. 

To calculate the background, we extract the counts from the stacked maps in an annular region within 50--70\arcsec\ of the central source. Since the PSF of the central sources are $\sim$ 2--3\arcsec\ the region for background extraction is, by construction, at a substantial distance from the sources. We find that changing the area of the region for background extraction does not affect the measured background as long as the region is sufficiently far from the sources. The background counts are estimated to be $1.68 \times 10^{-10}$ cm$^{-2}$s$^{-1}$pixel$^{-1}$, $3.4 \times 10^{-10}$ cm$^{-2}$s$^{-1}$pixel$^{-1}$, and $2.2 \times 10^{-10}$ cm$^{-2}$s$^{-1}$pixel$^{-1}$ respectively for the soft, hard, and ultra-hard bands.

\section{Results}  

Fig.\ 3 shows the total X-ray counts of our galaxies as a function of stellar mass. The left-hand, middle and right-hand panels are for the soft, hard, and ultra-hard bands. The total counts are calculated using an extraction scale of 3 arcseconds. Note that the redshifts of these sources are similar and hence the total counts are fair representations of their luminosities. To estimate errors we compute the standard deviation of counts within the extraction region and then divide by the square-root of the total number of pixels in the circle to obtain the standard error on the mean. Since we are interested in the total counts, we multiply the standard error by the total number of pixels. Hence the error-bars are $\sigma \sqrt N$, where $\sigma$ and N refer to the standard deviation of counts and the total number of pixels in the circle. 

It is important to note that according to this definition of our error on the counts, the signal-to-noise is proportional to $\sqrt N$. Hence a larger number of pixels within the extraction radius would result in a more accurate detection, if all other quantities stay the same. The open circles with arrows represent $2 \sigma$ upper limits of non-detected emission (i.e., the signal for these data points is consistent with the background to $2 \sigma$). The dashed line in each panel refers to the best-fit power law. We have used the weighted least square fitting method in this case \citep{A&H12}. The missing data points depict non-detections where the $2\sigma$ upper limits are either below the range shown in Fig.\ 3 or are consistent with zero counts.

We also note that in the current method of background extraction the actual amount of background remains uncertain since we do not know if a single photon came from the source or the background. So an uncertainty on the background should be considered while computing the error on the total counts from the source. As mentioned above we have extracted our background from a much larger region compared to the source extraction radius. In addition, we have checked that the background remains the same if we increase the size of the annulus. Due to the large area of the annulus from which we have extracted the background we note that the uncertainty on the background counts will be smaller than the Poisson noise we have considered. Hence we did not include the associated error in our error budget. 

In the soft band we see a clear correlation between X-ray counts and stellar mass. In the hard band (panel b of Fig.\ 3) we do not detect significant X-ray counts for the lower stellar mass bins. In the ultra-hard band (panel c of Fig.\ 3) our detections are less significant than in other bands. We used both significant and insignificant detections while computing the best-fit power-laws for Fig.\ 3. We note that the soft and the hard bands reflect similar slopes for the power-law. The slope in the ultra-hard band is consistent with zero. It is, thus likely that the bulk of the emission in the soft and the hard bands are coming from similar source population. While computing the total counts in Fig.\ 3 we have fixed the extraction radius for each stellar mass bin. However we note that since halo masses are correlated with stellar masses the sizes of galaxies and the corresponding dark matter halos will be different in each case. We thus compute the counts using scales corresponding to a fixed fraction $(f)$ of R$_{200}$ (defined as the radius at which the mean density of the halo is $200$ times the critical density of the Universe) derived from the respective halo masses. R$_{200}$ is computed from the dark matter masses using the theoretical scaling relation with stellar mass from \citet{mosteretal10}. We note that the best-fit slopes are statistically identical in each case.

From the counts in Fig.\ 3 we compute the hardness ratios. The hardness ratio is defined as $(H-S)/(H+S)$, where $H$ corresponds to the total counts in the hard band and $S$ refers to the total counts in the soft band. The hardness ratios are shown in Fig.\ 4. Note that for the non-detections we use the observed value of the counts and their observed errors. This accounts for the large error-bars in the hardness ratios in the lower stellar mass bins. To compare our hardness ratios with different spectral models of emission we compute model hardness ratios for a fixed flux ($10^{-16}$ ergs s$^{-1}$ at soft X-ray energies) with different spectral models using the web-based PIMMS tool.\footnote {http://heasarc.nasa.gov/Tools/w3pimms.html} 

The different models are a) power law with positive photon index, b) thermal black body, c) Bremsstrahlung, and d) the Astrophysical Plasma Emission Code (APEC; \citealt{smithetal01}). We assume a Galactic $N_{{\rm H}}$ of $1.3 \times 10^{20}{\rm cm}^{-2}$ \citep{d&l90} and a redshift of $z=0.8$. To study the effect of intrinsic absorption on the hardness ratios we calculated the model hardness ratios using an intrinsic $N_{H} = 5\times10^{-20} {\rm ergs}^{-1}$. We note that the change in hardness ratios due to absorption in some models is insignificant for the purpose of our analysis.

We find that in our definition of the hardness ratio, the pure black-body spectrum gives positive hardness ratios for $K_{B}T \ge 1.5 kev$. Our data thus exclude the pure black-body model for these energies. We compute the hardness ratios from the APEC model using different metallicities since the plasma emission will depend on metal abundances \citep{smithetal01}. We find that the differences in the hardness ratios due to differences in metallicities are well within our error-bars. So for our purpose we compute hardness ratios using solar abundances. In the top panel of Fig.\ 4 we plot the hardness ratios for the APEC model with solar metallicity as a function of plasma temperature. The Brehmsstrahlung model is depicted by blue triangles, as a function of temperature. In the bottom panel we plot the derived hardness ratios for a power-law model, as a function of photon index. The black squares correspond to the measured hardness ratios of our galaxy sources.

We note that X-ray emission from different sources will originate from different spatial scales in galaxy X-ray profiles, and ideally both spectral and spatial resolution of the sources will give us information about the exact source population. However, we do not have the spectral or spatial capability to distinguish the point sources in our high redshift galaxies from the extended emission. Recently, \citet{andersonetal13} performed a detailed stacking analysis of nearby isolated galaxies using ROSAT All Sky Survey data \citep{vogesetal99} to search for hot halos. At redshifts below $0.06$ their results show that the X-ray emission in isolated galaxies within a spatial scale of 50 kpc from the center of the galaxy is very likely to be a linear combination of a power-law emission from X-ray binaries and an APEC plasma emission from the hot halo. 

\citet{andersonetal13} performed detailed simulations to identify the degree of this degeneracy in isolated local galaxies and showed that the exact fraction of extended emission to power-law emission depends on the spatial scale and other parameters of the galaxy. We thus follow a simpler approach to compute the luminosities of our source galaxies. 

Following the arguments of \citet{andersonetal13} we extract the X-ray luminosities of our galaxies assuming an equal contribution from both extended emission and power-law emission. Our measured hardness ratios also suggest a similar level of degeneracy as seen by \citet{andersonetal13}, mostly due to the errors in our measured hardness ratios.\footnote{Note that our derived model hardness ratios are slightly different from \citet{andersonetal13} due to differences in the telescope response function and exact energy scales of the soft and hard bands.} We note that the assumption of Gaussianity in the error-propagation for small number counts can lead to a $30\%$ overestimate of our errors. A more accurate method involves the Bayesian Estimation of Hardness Ratios \citep{parketal06} approach.

\section{Discussions}

\begin{figure}[t]
\begin{center}
\begin{tabular}{c}
        \resizebox{8cm}{!}{\includegraphics{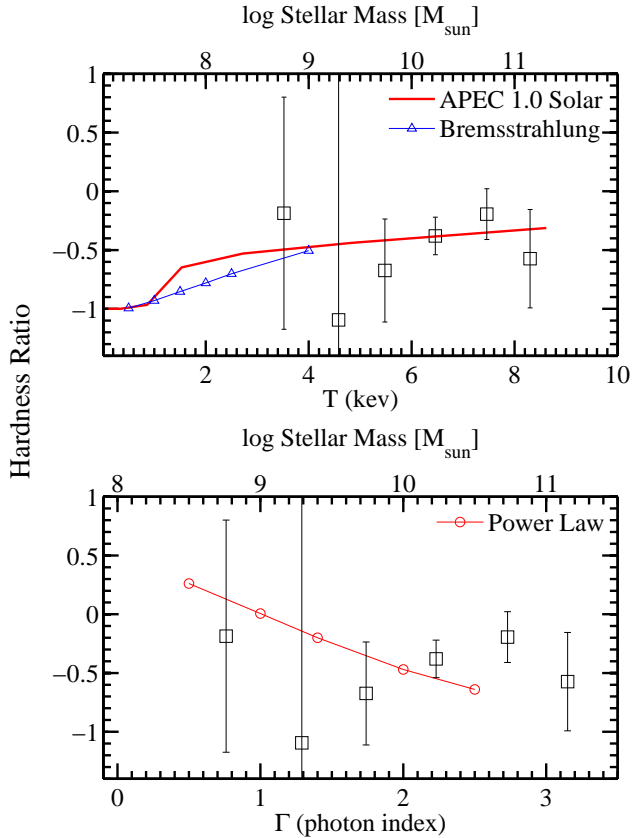}}
      
\end{tabular}
 \caption{Hardness ratios for different spectral models. The top panel shows the hardness ratios for the Bremsstrahlung model (blue solid line with triangles), and the APEC plasma model as a function of plasma temperature. In the bottom panel the red solid line with open circles shows the value of hardness ratios for a power law model with different photon indices. The black squares in each panel represent the measured hardness ratios of our sample as a function of stellar mass. The hardness ratios for the non-detections are calculated using the exact observed value of the counts and the exact observed error. That explains the large error-bars in the lower stellar mass bins. The hardness ratios seem to be degenerate between power law emission and APEC/Bremsstrahlung emission as is also observed in the case of low redshift galaxies by \citet{andersonetal13}. See \S 4 for more details.  }
\end{center}
\end{figure}

Previous studies have investigated the contribution from different kinds of sources to the X-ray luminosity of nearby galaxies, and their respective correlation with the stellar mass or the $K$-band or the $B$-band luminosity of the galaxy \citep[e.g.,][]{gilfanov04, b&g08, revnivtsevetal08, b&g11, borosonetal11}. Recent work \citep[][B11 hereafter]{borosonetal11} has shown that the correlation between X-ray luminosity and stellar population in a galaxy varies for different source populations. The different point sources are stellar X-ray sources (ABs and CVs), unresolved X-ray binaries (XRB), and resolved low mass X-ray binaries (LMXB). Thus the power-law emission in our high redshift galaxies will be the integrated emission from all of these sources. Also our data will possibly include some contribution from unresolved AGN, which have not been individually detected in our X-ray catalogs but can contribute to the signal in the co-added maps. From Fig.\ 4 we find that the typical photon index for our sources lies in the range 1.5--2.5. This spectral index falls in the range of both LMXBs \citep[e.g.,][]{grimmetal02, fabbiano06, burkeetal13}, and AGN sources \citep[e.g.,][]{n&p94, scottetal11, lanzuisietal13}.

It is well known that AGN activity is mostly seen in massive galaxies and feedback from AGN has been linked with suppression of star formation \citep[e.g.,][]{kauffmannetal03, dunlopetal03, schawinskietal07, alonsoherreroetal08, brusaetal09, vitaleetal13}, but the exact correlation between AGN feedback and galaxy stellar mass is yet to be confirmed. Some studies suggest that there is no correlation between stellar mass and AGN activity \citep[e.g.,][]{airdetal12}. Hence we assume that although nuclear X-ray emission from unresolved AGN can contaminate the X-ray emission observed at the high stellar mass end of Fig.\ 3 (see panel a), it will not be responsible for the emission observed in the entire stellar mass range. Thus we adopt a photon index of 1.5, typically assumed for emission from stellar sources in galaxies. We note that the supernova heated X-ray plasma in star-forming galaxies can have photon indices of 2. We stress that our data does not exclude such possibilities. However for the purpose of our luminosity calculations we note that a photon index of $1.5$ and $2.0$ does not affect our results significantly. We thus stress that our results could also represent the plasma in star-forming galaxies.     

To understand the ISM emission in our galaxies we need to have knowledge about their host dark matter halos. Recently \citet{mosteketal13} studied the clustering of DEEP2 galaxies at $z \sim 1$ as a function of stellar mass, finding that clustering strongly depends on stellar mass for blue star-forming galaxies in the mass range $9.5 < {\rm log(M_{\star}/M_{\odot})} < 11$, but has no dependence on stellar mass for red quiescent galaxies. The host dark matter halos of the blue galaxies are in the range $12.32 < {\rm log M}_{\rm halo}{\rm [h^{-1}M_{\odot}]} < 12.75$, whereas red galaxies in the stellar mass range $10.5 < {\rm log(M_{\star}/M_{\odot})} < 11.5$ are residing in host halos of masses $12.95 < {\rm log M}_{\rm halo}{\rm [h^{-1}M_{\odot}}] < 13.05$. In summary, clustering results suggest that galaxies with stellar mass $> 10^{10} {\rm h}^{-1}{\rm M}_\sun$ generally reside in massive galaxy/galaxy group scale halos. 

To check the distribution of galaxy colors in our sample we apply a high redshift color-cut, where we identify red galaxies with restframe color $(U-B) > 1$. We find that the fraction of red galaxies varies strongly with stellar mass. This, of course, is reflective of the selection criteria for DEEP2, where low mass galaxies are selected if they are bright enough in $B$-band. We find that roughly 30--40\% of our sources in the highest stellar mass bins are red galaxies. Thus, while computing the plasma emission for our galaxies, we adopt temperatures corresponding to massive galaxy scale halos ($\sim 10^{12.3}M_{\odot}$) for our low mass galaxies and group scale halos ($\sim 10^{13}M_{\odot}$) for high mass galaxies. From the two-component fitting (power-law emission with a photon index 1.5 plus a massive galaxy/galaxy group scale halo plasma emission) we obtain the soft X-ray fluxes of our galaxies and use the median redshift to convert to soft X-ray luminosities. The results are shown in Fig.\ 5.

\begin{figure}[t]
\begin{center}
\begin{tabular}{c}
        \resizebox{8cm}{!}{\includegraphics{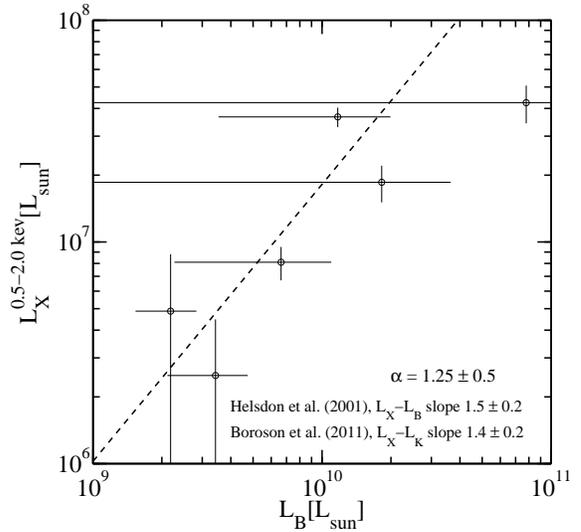}}
      
\end{tabular}
 \caption{$L_{X}-L_{B}$ relation for our high redshift galaxies. The X-ray luminosity has been calculated using a simple two-component spectral model where $50\%$ of the emission comes from the stellar sources in the galaxy (power-law emission) and the other $50\%$ is assumed to come from the hot ISM (APEC plasma emission; see Fig.\ 4 for the degeneracies in the hardness ratios between different spectral models). The spread in $B$-band luminosity is large due to our primarily mass-selected sample. The best-fit power-law slope is $1.25\pm 0. 5$, consistent with the results obtained for local galaxies using $B$-band luminosities (\citealt{helsdonetal01}), and $K$-band luminosities (\citealt{borosonetal11}).}
\end{center}
\end{figure}

We note that we are interested in the correlation between galaxy stellar mass and X-ray luminosity, but the majority of the low redshift studies use the $K$-band or the $B$-band luminosity of the galaxy to search for such correlations. We thus compute the average $B$-band luminosities of our stellar mass-selected sample and compare our results with local galaxies. We emphasize that the $K$-band luminosity is better correlated with stellar mass than $B$-band luminosity but our restframe $K$-band luminosities are incomplete mostly due to confusion in Spitzer data \citep{bundyetal06}. So, we rely on $B$-band luminosities. We note that our galaxies are selected based on stellar mass and thus in each mass bin we do observe a substantial scatter in B band luminosity. This should be taken into consideration while assessing the strength of the reported correlation. Another important effect that can alter the correlation is the morphological types of our galaxies (most of the low redshift scaling relations hold true for sample of {\em early-type} galaxies). We do not have information on galaxy morphology and hence we performed a spectral cut and re-examined our scaling relations based on this color-cut. As noted previously, the fraction of red galaxies varies strongly with stellar mass due to the selection criteria of the DEEP2 survey---neglecting that selection effect, the X-ray scaling relations for red galaxies are statistically identical to the X-ray scaling relations for the full sample.   

We now compare our results with the scaling relations for local galaxies. B11 showed that the correlation between X-ray luminosity and stellar $K$-band luminosity in a galaxy varies for different source populations. We do not have the spectral or spatial resolution to individually resolve the components that are responsible for the source population. We will thus compare our total X-ray luminosity (which we have obtained assuming an equal contribution from the power-law emission and extended emission) slopes with B11 and other studies that used local samples. B11 argue that for LMXBs the slope of the power law is $\sim 1$ as found in previous studies \citep[e.g.][]{whitereetal02,k&f04,davidetal06}. For the diffuse hot gas the slope is steeper compared to the LMXBs (B11). There has been debate in the literature regarding the scatter of the $L_{X} - L_{K}$ relation for hot gas. This difference in slope could arise due to differences in the dark matter potential well (or the environment) of the galaxy \citep[e.g.][]{jeltemaetal08, m&j10}. B11 attribute another component as a contributor to the X-ray emission in galaxies; namely unresolved LMXBs (designated ``unresolved nuclear emission'' in B11). For this particular component B11 do not find any correlation between X-ray luminosity and $K$-band luminosity. 

The $L_{X} - L_{K}$ correlations for different X-ray sources in low redshift galaxies are summarized in Fig.\ 2 of B11. Similar results should also be expected for $B$-band luminosities (B11). Our best-fit slope for the  $L_{X} - L_{B}$ scaling shown in Fig.\ 5 is $1.25 \pm 0.5$, consistent with the total $L_{X} - L_{K}$ slope of B11 ($1.4 \pm 0.2$) and the $L_{X} -L_{B}$ slope of \citeauthor{helsdonetal01} (\citeyear{helsdonetal01}; $1.5 \pm 0.2$) and \citeauthor{shapleyetal01} (\citeyear{shapleyetal01}; $1.5 \pm 0.1$). Although this is consistent with our findings we emphasize that our scaling relations with luminosity are likely to be affected by both selection biases and theoretical biases (e.g.\ assumption of the nature of the source populations) as discussed previously.

\section{summary}
In this study we have detected significant X-ray emission from normal galaxies at high redshift ($\sim$0.8) as a function of stellar mass. We have shown that there exists a significant correlation between stellar mass and X-ray emission at scales within 10\,kpc in the soft X-ray band in these high redshift galaxies. We derived a power-law relationship between integrated X-ray counts and galaxy stellar mass. The slope of the best-fit power law is $0.58 \pm 0.1$. We compute the hardness ratios of our sources and show that the hardness ratios of our sources are degenerate between two spectral models, namely point-like power-law emission and extended plasma emission in the ISM. This has been robustly demonstrated by \citet{andersonetal13} in the case of nearby galaxies and our results are similar to those observations. 

Using a simple two-component spectral model where half of the emission comes from hot halos and the other half comes from stellar point sources in the galaxy, we compute the approximate luminosities ($10^{39} < L_{X}^{{\rm soft}} < 10^{41}$ ergs s$^{-1}$) of our sources. We also compute the $B$-band luminosities of our sources and show that the $L_{X}-L_{B}$ scaling relation is consistent with low-redshift results (B11). We also made a color cut $(U-B) > 1$ to our galaxy sample to examine the effect of galaxy color (broadly equivalent with morphology) on the scaling relations. The scaling relations are statistically identical when such a color-cut is imposed, within the limitations of the DEEP2 survey selection. Deeper extragalactic X-ray surveys should provide additional insights into the nature of the X-ray source populations in high redshift galaxies, which can reveal important clues about galaxy evolution. Our analysis provides a first step in this direction.         

\section*{Acknowledgments}
SC thanks Daisuke Nagai for very important discussions regarding the initial development of the project. We also thank the referee for providing several important suggestions that were useful in improvement of the paper. SC and ADM were partially supported by the National Science Foundation through grant number 1211112, and by NASA through ADAP award NNX12AE38G and Chandra Award Number AR0-11018C issued by the Chandra X-ray Observatory Center, which is operated by the Smithsonian Astrophysical Observatory for and on behalf of NASA under contract NAS8-03060. SC was partially supported by NSF grant AST-0806732 at the University of Pittsburgh during the course of this work. 

\bibliography{mybib}{}
\bibliographystyle{apj}
\end{document}